\documentstyle[12pt,aaspp4]{article}
\def\gtorder{\mathrel{\raise.3ex\hbox{$>$}\mkern-14mu
                \lower0.6ex\hbox{$\sim$}}}
\def\ltorder{\mathrel{\raise.3ex\hbox{$<$}\mkern-14mu
                \lower0.6ex\hbox{$\sim$}}}
\def\eg{{\it e.g.~}}
\def\etal{{\it et al.~}}

\lefthead{}
\righthead{}
 
\def\lta{{\>\rlap{\raise2pt\hbox{$<$}}\lower3pt\hbox{$\sim$}\>}}
\def\gta{{\>\rlap{\raise2pt\hbox{$>$}}\lower3pt\hbox{$\sim$}\>}}

\begin{document}

\title{Large stellar disks in small elliptical galaxies}

\author{Hans-Walter Rix$^{1,4}$, Marcella Carollo$^{2,5}$ and Ken Freeman$^3$}

\vfill\noindent
Astrophysical Journal Letters, March 1999.
\bigskip

\noindent $^1$ Steward Observatory\\
$^2$ Johns Hopkins University\\
$^3$ Mount Stromlo and Sidings Springs Observatory\\
$^4$ Alfred P. Sloan Fellow\\
$^5$ Hubble Fellow\\

\begin{abstract}
We present the rotation velocities $V$ and velocity
dispersions $\sigma$ along the principal axes of seven elliptical
galaxies less luminous than 
$M_B= -19.5$.  These kinematics extend beyond the
half-light radii for all systems in this photometrically selected
sample. At large radii the kinematics not
only confirm that rotation and ``diskiness"  are important in faint ellipticals, as
was previously known, but also demonstrate that in most sample galaxies the stars  at large
galactocentric distances have $\bigl (V/\sigma\bigr )_{max}\sim 2$, similar to the
disks in {\it bona-fide}
S0 galaxies.  Comparing this high degree of ordered stellar motion
in all sample galaxies
with numerical simulations of dissipationless mergers
argues against mergers with mass
ratios $\le 3:1$ as an important mechanism in the final shaping of
low-luminosity ellipticals, and favors instead the dissipative
formation of a disk. 
\end{abstract}

{\it subject headings}: galaxies: formation - galaxies: evolution -
galaxies: structure - galaxies: elliptical and lenticular - galaxies: kinematics and dynamics

\section{Introduction}

\noindent

\noindent
Elliptical galaxies occupy only a small volume in the stable 
 parameter space defined by mass, luminosity, compactness, and
rotational support (\eg Kormendy and Djorgovski 1989; de~Zeeuw and Franx 1991).
This means that their present-day structure is not
determined by stability, but by their formation history, specifically by the relative
importance and the time ordering of dissipative processes {\it vs.} merging. 
Yet, the actual role of gas dissipation, resulting in gaseous and stellar disks, 
and of violent relaxation, leading to
spheroidal systems with largely random stellar motions, is still
under debate for ellipticals of different luminosity classes.

\noindent
Observationally it has been established that ``low-luminosity" ellipticals 
($L\ltorder \frac{1}{2} L_*$) show more rotation and photometric
diskiness than the brighter systems (e.g., Davies \etal 1983; Bender
et al.\ 1989).  Yet, taken alone, these results cannot constrain uniquely
their formation history, as there are many paths towards
systems with modest rotation. For example,  N-body
simulations of dissipationless major mergers can produce rotating
remnants $(V/\sigma)_{max}\ltorder 1$ 
with disky isophotes (e.g., Heyl, Hernquist \& Spergel 1996;
Weil \& Hernquist 1996).  On the basis of statistics, Rix \& White
(1990) argued that most of the so-called faint ellipticals are not
just ``disky'' objects, but are rather the face-on counterparts
of S0 galaxies, implying  that they contain dynamically-cool stellar disks
that make up an appreciable fraction of the total mass.  If confirmed kinematically,
the presence of an outer stellar disk in the majority of these systems would strongly
support gas dissipation as the last significant step in their
formation history. For individual objects, detailed photometric and
kinematic studies have shown indeed that dynamically fragile stellar
disks are present in several elliptical galaxies (\eg Rix \& White 1992; 
 Scorza \& Bender 1995).

\noindent
In this paper we attempt to move beyond questions of individual
misclassification of galaxies, by asking whether there is
any significant fraction of ``faint elliptical galaxies'' that have
reached their present state through largely
dissipationless merging as the last formation step?  To this
end we present  and discuss kinematic measurements (to $R > R_e$) for a
photometrically selected sample of seven bona-fide ellipticals with $M_B \ge
-19.5$. 

\section{Sample and Observations}

\noindent
Our seven galaxies are a random sub-sample of the RC3 catalog (de
Vaucouleurs et al.\ 1991), with Hubble types $T\le -4$ and 
luminosities below $M_B= -19.5$.  No prior kinematic information was used,
and nothing, save the luminosity cut, should have biased the selection process
towards rapidly rotating objects. We added three S0 galaxies ($T =
-2,-3$) for comparison.  The spectra were obtained
during two different runs (February 11-14, 1997, and September 29 --
October 2, 1997) at the KPNO 4-m telescope, using the RC spectrograph
with the KPC-24 grating in second order.  In February 1997 the
detector was T2KB CCD with $2048^2$ pixels of $24\mu$m$^2$, which were
binned along the slit to yield $1.38''$/pixel.  In the September 1997 run we
used a 3k$\times$1k F3KB CCD with $15\mu$m$^2$ pixels binned to
$0.86''$/pixel along the slit.  With a $2.5''$ wide slit and on-chip
binning by two in the spectral direction the effective instrumental
resolution was $\sigma_{instr}\approx 50$~km/s.  The spectra were
centered on 5150$\AA$, covering H$\beta$ $\lambda$4861$\AA$, [OIII]
$\lambda$5007$\AA$, Mg$_b$ $\lambda$5175$\AA$, Fe $\lambda$5270$\AA$
and Fe $\lambda$5335.  Table 1 lists relevant properties of the sample members
and the observations, including the adopted major axis
position angle and the exposure times.
 Spectra of several K giants were acquired with the same
instrumental setup, and used as kinematic templates. 
The basic data reduction
(bias and dark subtraction, flat-fielding, correction for
slit-vignetting, wavelength calibration, airmass and Galactic
extinction corrections, correction for instrumental response) was
performed using {\tt IRAF} and {\tt MIDAS} software.

\noindent
The stellar kinematics were obtained with the template fitting method,
described in Rix \& White (1992), after the data were
binned along the slit to constant
signal-to-noise. Then the best fit mean velocity $V$ and dispersion
$\sigma$ were determined by $\chi^2$-minimization, along with the best
fitting composite stellar template. The resulting error bars
are the formal uncertainties based on the known sources of
noise. External tests on this and other data sets demonstrate that
they are a good approximation to the true uncertainties. 

\section{Results and Discussion}

\noindent
Figure 1 shows the major and minor axis kinematics for the sample
galaxies.  Similar to S0 galaxies, the velocity dispersions in these
ellipticals are found to drop from a central peak to near our
instrumental resolution at large radii.  Most sample
members exhibit little - if any - rotation along the minor axis, in contrast
to their strong rotation along the major axis.  Their kinematics
suggest that  these systems are nearly axisymmetric and likely
have a rather simple dynamical structure.  Only NGC1588 shows considerable
minor axis rotation over the
same radial region where the photometry shows isophotal twisting; this is
possibly due to the interaction with its close companion (NGC1589).

In Figure 2 we quantify the degree of rotational support in these
galaxies, by plotting $V/\sigma$ as a function of the galactocentric
distance along the major  axis.  As a diagnostic of the dynamical state
beyond the effective radius, we adopt the maximum, $(V/\sigma)_{max}$,
and the outermost value $(V/\sigma)_{out}$, of each $V/\sigma$ curve
[see Table 1].  
 In terms of these quantities, Figure 3 re-states the main
result: for four out of seven ellipticals, $(V/\sigma)_{max}\gta2$;
two additional ellipticals have $(V/\sigma)_{max}\gta1.5$.

As a benchmark, we compare the observed $(V/\sigma)_{max}$ [and
$(V/\sigma)_{out}$] values to 
maximally rotating, oblate ``Jeans models.''  We consider models where the
residual velocities are isotropic in the co-rotating frame. 
The models are built following the method outlined in Binney, Davies, \&
Illingworth (1990; see also Carollo \&
Danziger 1994 for particulars),
assuming a Jaffe (1983) law.
The curves in Figure 3 represent maximum $V/\sigma$
within $2 R_e$, as a function of projected axis ratio (or
inclination). We considered {\it intrinsic} axis-ratios of $c/a=0.6$ and
$c/a=0.4$ and
models with and without dark halos (which are three times more
extended and more massive than the luminous component).  As Figure 3
illustrates, no model can reproduce values of $(V/\sigma)_{max}\sim 2$ 
 for any viewing angle, unless it has $c/a\ltorder 0.4$! This
implies that all sample galaxies are intrinsically {\it very} flat. A
possible exception is the one distorted galaxy, NGC1588, for which our kinematics are compatible
with $c/a<0.6$, if $(V/\sigma)_{out}$ is considered instead of
$(V/\sigma)_{max}$. The probability that 6 out of 7 ellipticals
 are that flat is vanishing ($\approx 2.4 \times 10^{-6}$)
on the basis of the expected shape distribution
of low-luminosity oblate spheroidals (Tremblay \&
Merritt 1996).

Dissipationless simulations of equal mass galaxy mergers 
mostly produce  remnants with  $(V/\sigma)_{max} \ltorder 1$ 
inside $\sim 2R_e$ (\eg Heyl, Hernquist
and Spergel, 1996).   In addition, for most spin-orbit geometries major mergers
also produce significant kinematic misalignment, which would be reflected in
minor axis rotation.  Neither property is consistent
with our sample galaxies, ruling out formation through mergers of nearly equal mass.
Unequal mass mergers ($\sim 3:1$) can produce remnants with rather disk-like
shapes and kinematics (Barnes, 1996; Bekki 1998; Barnes 1998). Barnes
(1998) finds remnants to be close to axisymmetric with little minor axis rotation,
in this respect
consistent with our sample properties. 
He quantifies the rotational support of the merger remnants by the 
parameter $\lambda'$, the total angular momentum of the most tightly
bound half of the stars, normalized by the value for
 perfect spin alignement.
 For Barnes' eight  3:1 mergers $\lambda'_{sim}=0.38$ with a scatter of $ 0.04$.
As $\lambda'_{sim}$ is not observable, we attempt to construct an analogous 
quantity $\langle\lambda'\rangle_{obs}$ for our sample members
(see Table 1), by estimating the azimuthal velocity, normalized by the local
circular velocity and averaged  over the inner
50\%  of the light. For this estimate we had to make the following
assumptions: (i) the overall potential is logarithmic; (ii) the velocity
dispersions are isotropic; (iii) the galaxy is
axisymmetric with an intrinsic axis ratio of $0.4$ (or $0.6$; see above); (iv)
the circular velocity is estimated as $v_c\approx \sqrt{v_\phi^2+2 \sigma^2}$, which is
both correct for non-rotating systems in logarithmic potentials and in the
``asymmetric drift" limit for $\rho_*\propto r^{-2}$; (v) the stars are on average 
30$^\circ$ from the mid-plane.  With this, we find
$\langle\lambda'\rangle_{obs}=0.55$ with a scatter of $0.06$ (see
Table 1): 
every observed sample galaxy has  considerably more ordered motions than
any of  the 3:1 mergers. We checked that plausible changes
in  the above model assumptions
will alter not $\langle\lambda'\rangle_{obs}$ by 
$\langle\lambda'\rangle_{obs} - \lambda'_{sim}$.

The degree of streaming motion found in our sample galaxies is nearly that of the
dominant disks of S0 galaxies [\eg Fisher 1997 and the
$(V/\sigma)_{max}$ values for the S0s of our sample].  
Without exception the sample members appear even more rotationally supported than
simulated unequal mass (3:1) mergers (Barnes 1998).
This strongly suggests that the dissipative
formation of a massive and extended stellar disk has been the last major step
in building these galaxies. This inference does not preclude, however, that for
some fraction this disk has been heated considerably by subsequent gravitational interactions.

Our data, therefore allow us to push the long-known result that
{\it rotation is important in low-luminosity ellipticals} (Davies \etal 1982) one step
further: most of these ``elliptical" galaxies in an effectively luminosity-limited sample,
actually contain stellar disks at large radii, comparable in mass and size to S0s.
These kinematic results, combined with earlier
photometric evidence ( e.g., Rix \& White 1990 and references therein),
now put this idea on solid observational grounds. Perhaps these results imply
that at smaller mass scales the epoch of mergers ended before the epoch
of star-formation.

\bigskip
\bigskip
 
\acknowledgments

HWR is supported by the Alfred P. Sloan Foundation. 
CMC is supported by NASA through the grant HF-1079.01-96a awarded by
the Space Telescope Institute, which is operated by the Association of
Universities for Research in Astronomy, Inc., for NASA under contract
NAS 5-26555.  

\newpage

\begin{figure}
\epsfxsize=\hsize\epsffile{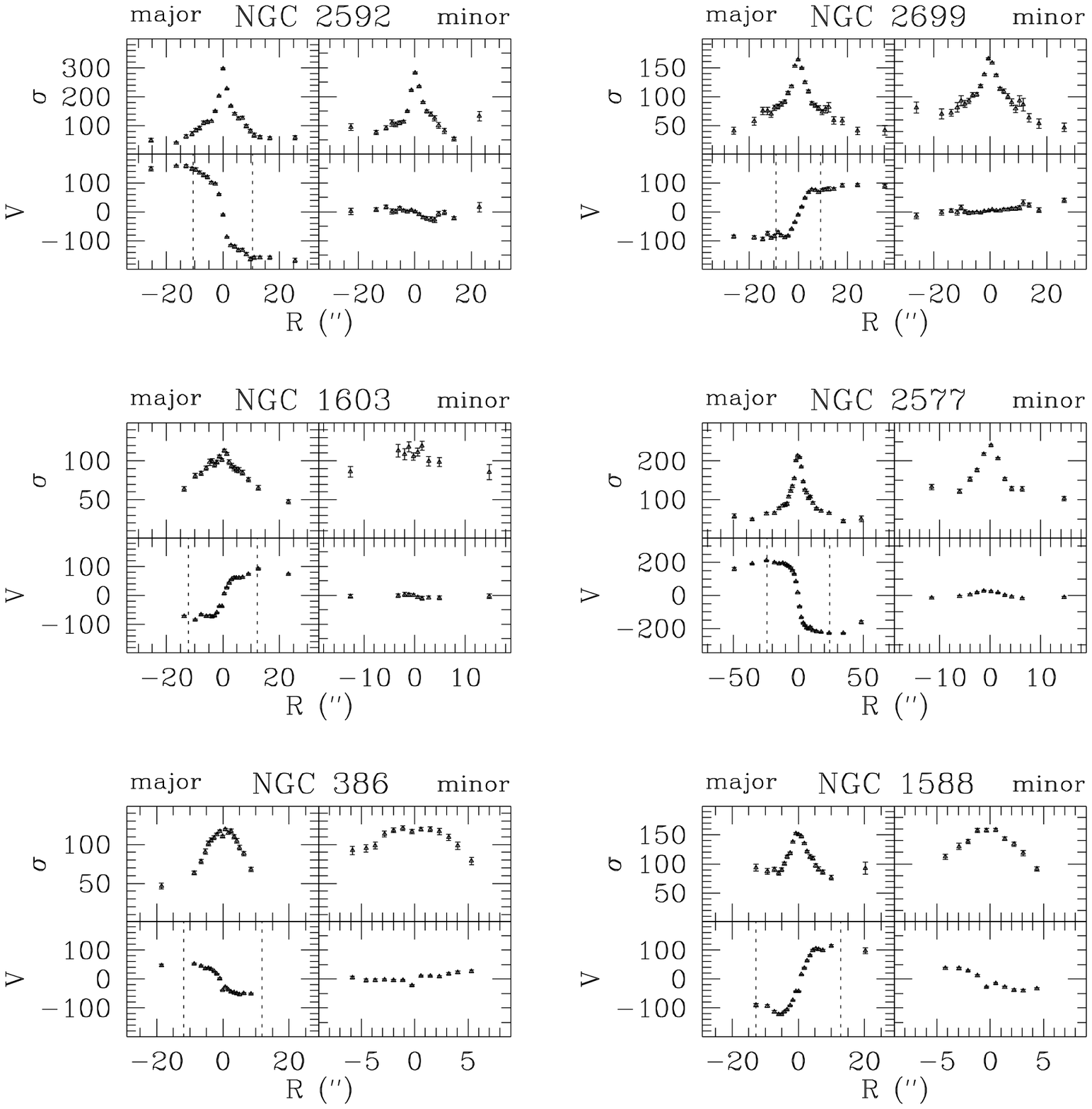}
\caption{Rotation velocity $V$ and velocity dispersion $\sigma$ (both
in km/s) along the principal axes as a function of radius (in seconds of arc) for the sample
galaxies. Vertical dotted lines are plotted at the half-light radii.}
\end{figure}

\begin{figure}
\epsscale{1.00}
\epsfxsize=\hsize\epsffile{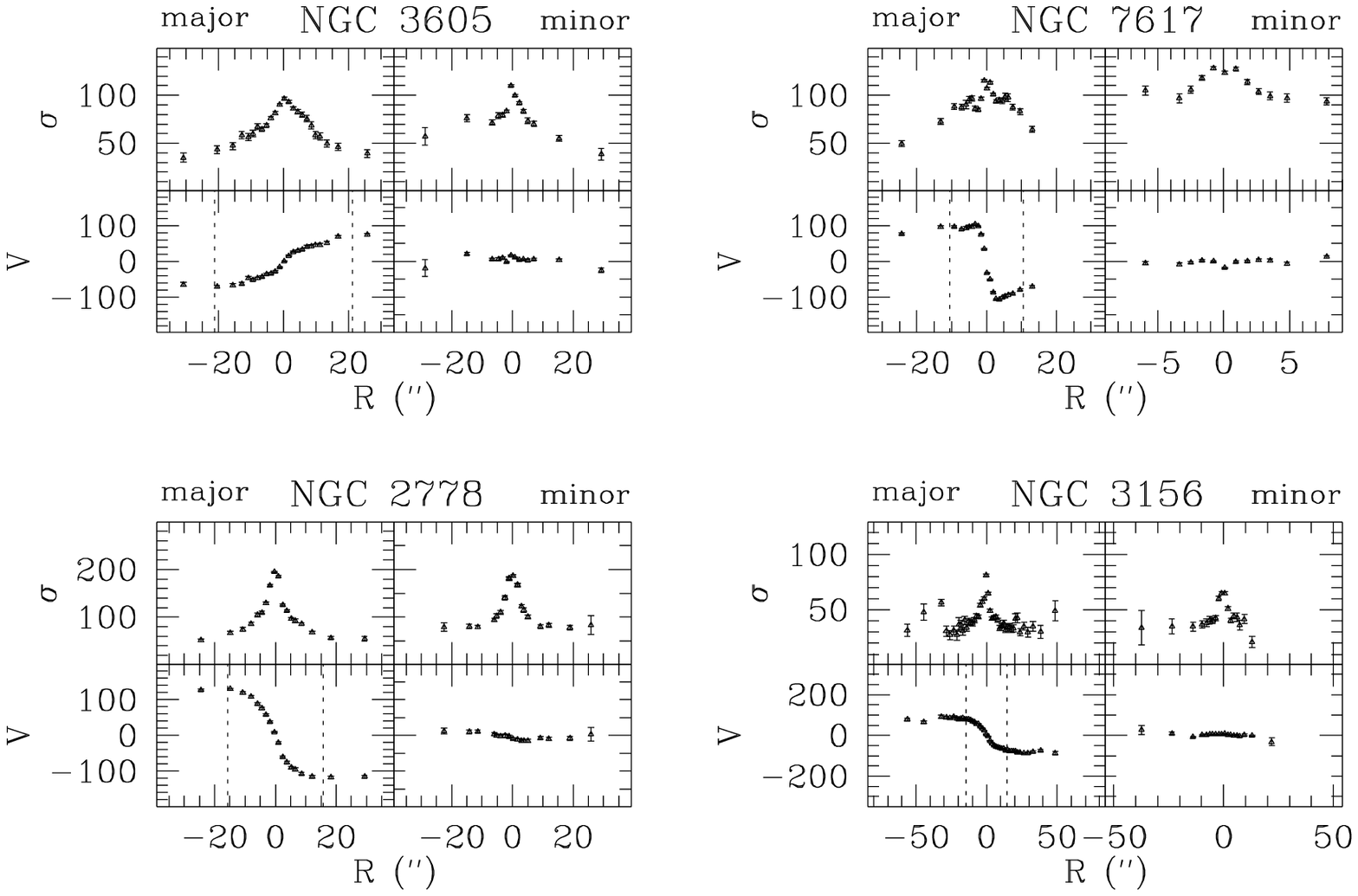}
\end{figure}

\begin{figure}
\plotone{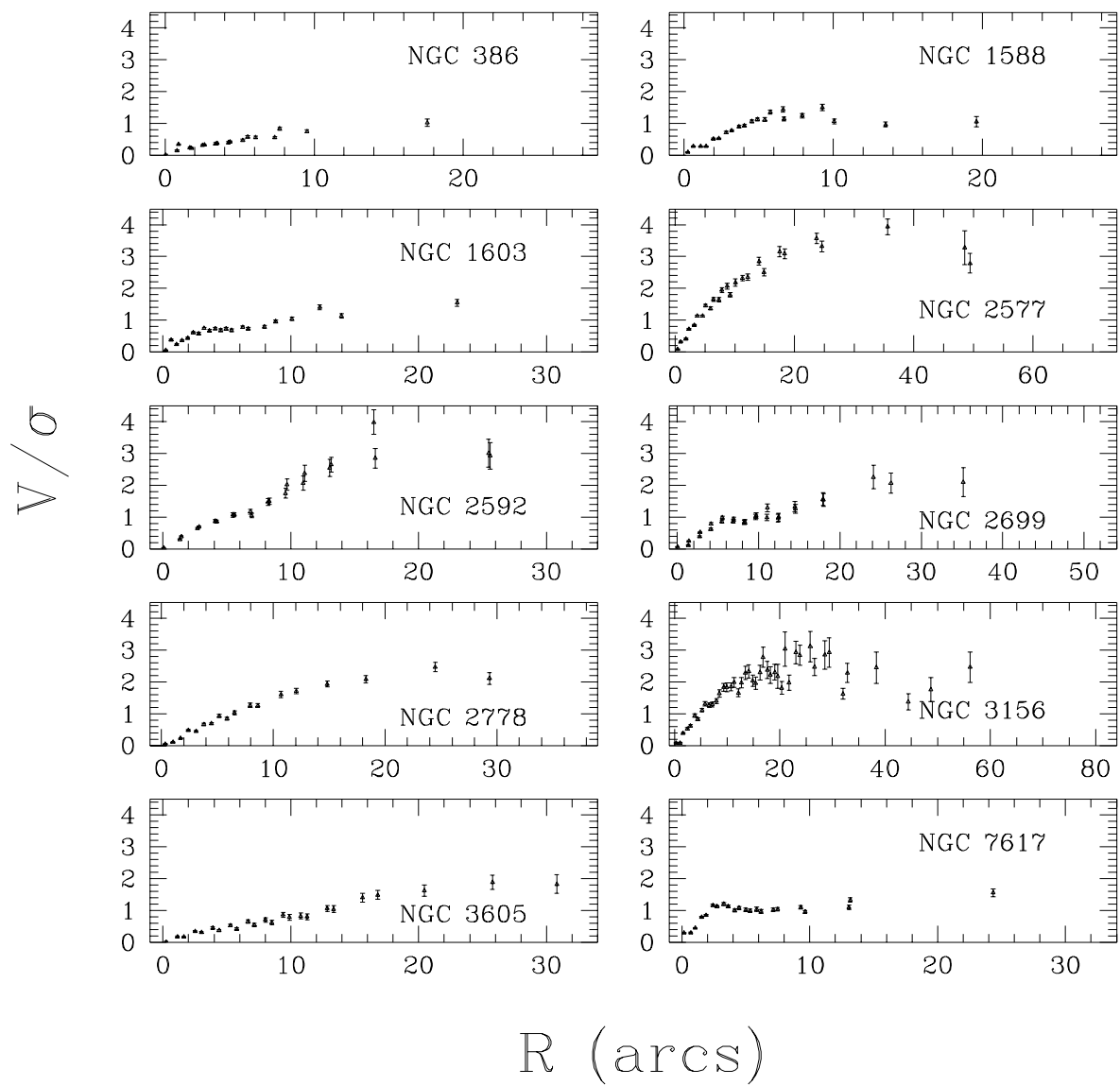}
\caption{The ratio of rotation velocity and velocity dispersion,
$V/\sigma$, as a function of radius (in arcseconds), illustrating the strong
outward increase in rotation support for almost all sample members.}
\end{figure}

\begin{figure}
\plotone{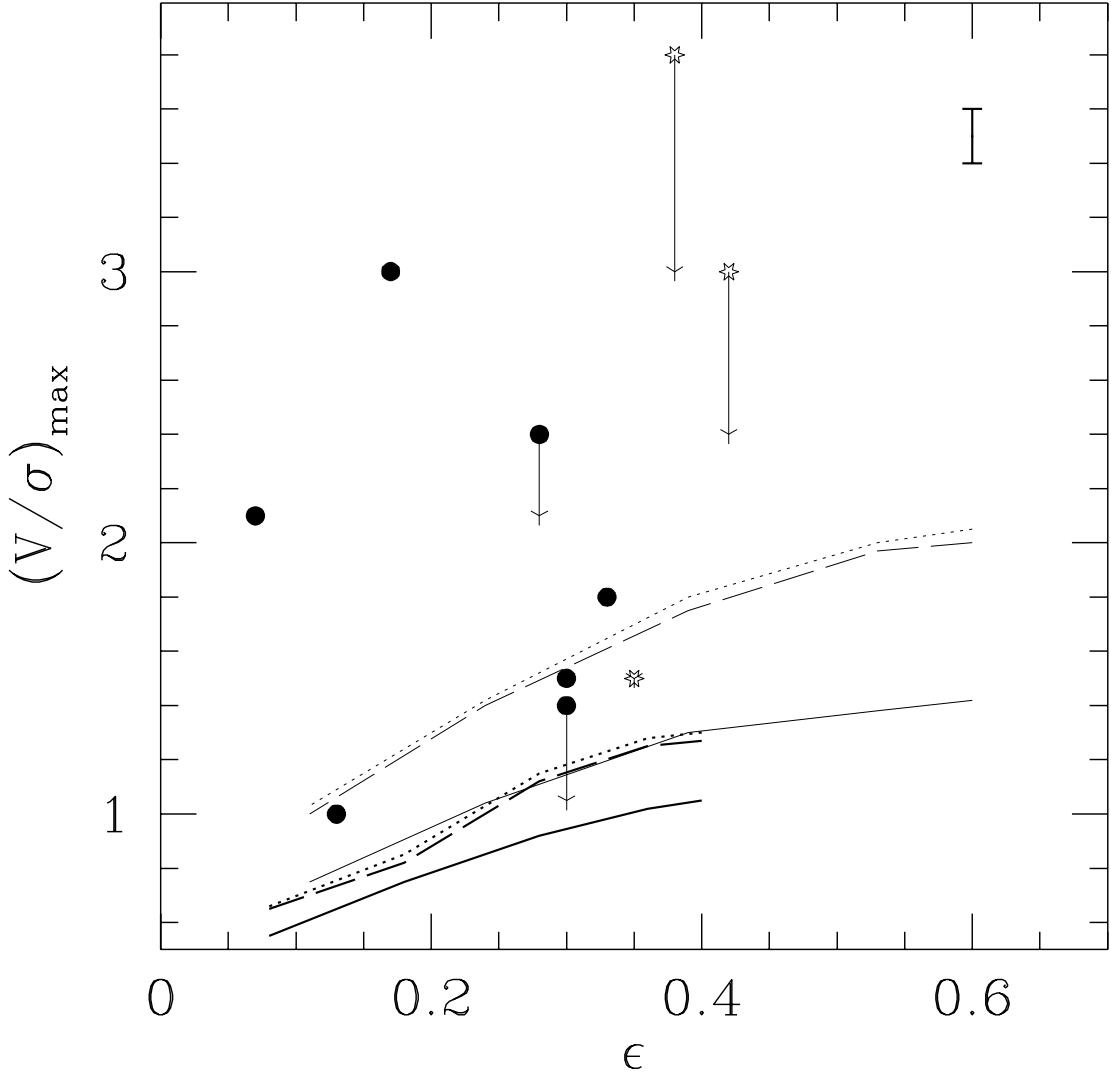}
\caption{Comparison of $(V/\sigma)_{max}$ (filled squares for the
ellipticals; stars for the lenticular galaxies) with oblate,
$f(E,J_z)$ Jeans models (thick lines for an intrinsic $c/a=0.6$, and 
thin lines for an intrinsic $c/a=0.4$).  The dotted lines indicate models with
no dark
matter; dashed lines represent models with round dark halos; the solid
lines indicate dark halos as flattened as the luminous component.  All
halos are three times more extended and more massive than the luminous
component. Arrows point to the corresponding values of
$(V/\sigma)_{out}$, when these differ from $(V/\sigma)_{max}$. The
typical errorbar on $(V/\sigma)_{max}$ [and $(V/\sigma)_{out}$] is
given in the upper-right corner.}
\end{figure}
\newpage

 \begin{table*}
\begin{center}
{\tiny\begin{tabular}{llccccccccccc}
\hline\hline
Name & $T$ & V$_{hel}$& $D$ & $M_B$ & $R_e$&$\epsilon$ & Run & T$_{maj/min}$
&$PA$ &$(V/\sigma)_{out}$ & $(V/\sigma)_{max}$ &$ \langle{\lambda'}\rangle$\\
     &    &  (km/s) & (Mpc) &  (mag) & (arcs) &&  & (secs) & (deg)& & & \\
\hline\hline  
NGC386&  -5  & 5707 & 76            & -19.2  &11.8$\ddagger$ &0.13   & Sep97&18000/8100& 5 &1.0&1.0&0.47 (0.43)\\
NGC1588& -4.6  & 3397 & 19$^\dagger$  & -17.7& 12.8 &0.30& Sep97  & 10800/3400 &35&1.05&1.4&0.58(0.54)\\
NGC1603& -5?  & 5038 & 67            & -19.5 &12.2  &0.30 & Sep97  & 18000/2100& 50&1.5&1.5&0.50 (0.46)\\
NGC2592& -5  & 1925 &  26           & -18.9  &10.5$\ddagger$  &0.17  & Feb97  &5400/5400 & 53& 3.0&3.0&0.65 (0.62)\\
NGC2699& -5  & 1660 &  22           & -18.1  & 9.0$\ddagger$ &0.07&  Feb97  &5400/5400 & 45& 2.1& 2.1&0.63 (0.60)\\
NGC2778& -5  & 1991 &  27           & -18.8  & 15.7 &0.28 &  Feb97  &5400/3600&40 &2.1&2.4&0.54 (0.50)\\
NGC3605& -5  & 581  & 14$^\dagger$  & -17.6  &  21.2&0.33&  Feb97  & 4400/5400&22 &1.8&1.8&0.47 (0.43)\\
\hline
NGC2577& -3  & 2068 &  28           & -19.0  & 24.3$\ddagger$&0.38 & Feb97  &5400/3600&105 & 3.0&3.8&0.71 (0.68)\\
NGC3156& -2  & 980  & 15$^\dagger$  & -17.9  & 14.4   &0.42 & Feb97  & 5400/5400& 47&2.4 &3.0&0.60 (0.57)\\
NGC7617& -2  & 4176 &  56           & -19.1  & 10.6  &0.35  & Sep97  &18000/12560 & 30& 1.5&1.5&0.62 (0.58)\\
\hline\hline
\end{tabular}
}
\vskip 0.5truein
\caption{Parameters and observing log for the seven elliptical and
three S0 galaxies of our sample.  Morphological types $T$, magnitudes
and (outer) ellipticities $\epsilon$ are from the RC3 or UGC
Catalogs. The half-light radius $R_e$ is from the RC3 catalog, except
for the galaxies identified with the symbol $\ddagger$, for which this
measurement is derived from our spectral data (since it is not
provided by RC3). Absolute magnitudes are corrected for Galactic
extinction. We assume $H_o=75$ km/s/Mpc.  Distances are computed
directly from the Hubble flow velocities (RC3), except for those
galaxies identified by the $\dagger$. For these, the peculiar motion
corrected distances from Faber et al.\ (1989) or Bender et al.\ (1992)
are adopted.  T$_{maj/min}$ gives the major and minor axis spectra
total exposure times, respectively. $PA$ gives the adopted values for
the position angle of the major axis. The columns $(V/\sigma)_{out}$
and $(V/\sigma)_{max}$ list respectively the outermost and the maximum
value of the $V/\sigma$ ratio over the radial range sampled by the
data. The spin parameter $\langle\lambda\rangle$ (for $q_{intr}=0.4$) is defined in the text;
values in parentheses refer to $q_{intr}=0.6$.}
\label{tab1}
\end{center}
\end{table*}
\normalsize

\end{document}